\documentclass[aip,pop,reprint,groupedaddress, amsmath, amssymb, showpacs]{revtex4-2}

 \usepackage{gensymb}
 \usepackage{subcaption}
 \usepackage{caption}
\newcommand*{\rom}[1]{\expandafter\@slowromancap\romannumeral #1@}
\newcommand{\vc}{\mathbf}
\usepackage{graphicx}
\usepackage{color}

\begin{document}

\title{A Kinetic Model of Friction in Strongly Coupled Strongly Magnetized Plasmas}


\author{Louis Jose}
\affiliation{Nuclear Engineering \& Radiological Sciences, University of Michigan, Ann Arbor, Michigan 48109, USA}

\author{Scott D. Baalrud}
\email{baalrud@umich.edu} 
\affiliation{Nuclear Engineering \& Radiological Sciences, University of Michigan, Ann Arbor, Michigan 48109, USA}


\date{\today}

\begin{abstract}

Plasmas that are strongly magnetized in the sense that the gyrofrequency exceeds the plasma frequency exhibit novel transport properties that are not well understood. As a representative example, we compute the friction force acting on a massive test charge moving through a strongly coupled and strongly magnetized one-component plasma using a generalized Boltzmann kinetic theory. Recent works studying the weakly coupled regime have shown that strong magnetization leads to a transverse component of the friction force that is perpendicular to both the Lorentz force and velocity of the test charge; in addition to the stopping power component aligned antiparallel to the velocity. Recent molecular dynamics simulations have also shown that strong Coulomb coupling in addition to strong magnetization gives rise to a third component of the friction force in the direction of the Lorentz force. Here, we show that the generalized Boltzmann kinetic theory captures these effects, and generally agrees well with the molecular dynamics simulations over a broad range of Coulomb coupling and magnetization strength regimes. The theory is also used to show that a "gyro" component of the friction in the direction of the Lorentz force arises due to asymmetries associated with gyromotion during short-range collisions. Computing the average motion of the test charge through the background plasma, the transverse force is found to strongly influence the trajectory by changing the gyroradius and the gyro friction force is found to slightly change the gyrofrequency of the test charge resulting in a phase shift.

\end{abstract}




\maketitle





\section{Introduction}


 
 Standard kinetic theories are designed to treat plasmas that are both weakly coupled and weakly magnetized. 
 Weakly coupled plasmas are a dilute gas like state in which the average potential energy of Coulomb interactions is small compared the average kinetic energy: $\Gamma \ll 1$, where $\Gamma = (e^2/a)/(k_BT)$ is the Coulomb coupling parameter and $a=[3/(4\pi n)]^{1/3} $ is the Wigner-Seitz radius. 
Weakly magnetized plasmas are characterized by the condition that particles gyrate at a scale that is much larger than the scale at which Coulomb interactions occur (Debye length), which implies that $\beta \ll 1$, where $\omega_c=eB/mc$ is the electron gyrofrequency and $\omega_p = \sqrt{4\pi e^2n/m}$ is the electron plasma frequency. 
 Here, we develop a generalized plasma kinetic theory that applies across Coulomb coupling and magnetization strength regimes, spanning from large to small values of both $\Gamma$ and $\beta$. 
 Exploring the combination of strong Coulomb coupling and strong magnetization is interesting from a fundamental physics perspective, but it is also important from a practical perspective as the plasma in many types of experiments can simultaneously span strong coupling and strong magnetiation regimes. 
 These include non-neutral plasmas~\cite{Beck_PRL_nonneutral}, ultra-cold neutral plasmas~\cite{Zhang_Ultracold_plasma}, anti-matter traps~\cite{fajans2020plasma_antimatter_traps}, electron cooling devices~\cite{Men_shikov_2008}, fusion experiments ~\cite{Aymar_2002} and magnetized dusty plasmas~\cite{Thomas_dusty_plasmas}. 
 Strongly coupled and strongly magnetized plasmas also occur naturally in astrophysical plasmas, such as the atmosphere of neutron stars and magnetars~\cite{Harding_2006}. 
 

Our approach combines the recent~\citep{Jose_POP_GCO} ``generalized collision operator'' that extends the Boltzmann equation~\cite{ferziger1972mathematical} to treat strong magnetization with the  mean force kinetic theory~\cite{Baalrud_PRL,baalrud2014extending} that extends the Boltzmann equation to treat strong coupling. 
The generalized collision operator for strong magnetization is similar to the Boltzmann equation, but where binary collisions occurring within a collision volume are computed numerically including the Lorentz force, which replaces a scattering cross section in the theory. 
The mean force kinetic theory is derived based on an expansion related to the deviation of correlations from their equilibrium values, rather than in terms of the strength of correlations. 
The result is similar to the Boltzmann equation, but where binary collisions occur via the potential of mean force, rather than the bare (Coulomb) potential. 
Here, we combine these two approaches to obtain a kinetic equation capable of spanning coupling and magnetization strength regimes. 

Other approaches have been proposed to extend plasma kinetic theory to strong magnetization or strong coupling, but not both. 
In order to appreciate the parameter regimes in which each of these is applicable, it is helpful to consider the coupling-magnitization parameter space proposed in Ref.~\onlinecite{Baalrud_PRE_2017}. 
At weak coupling $\Gamma < 1$, four magnetization regimes can be defined by comparing the gyroradius ($r_c = \sqrt{k_B T/m}/\omega_c $) with the Coulomb collision mean free path ($\lambda_{\textrm{col}}$), the Debye length ($\lambda_D = \sqrt{k_B T/ 4 \pi e^2 n} $), and the Landau length, which is the average distance of closest approach in a binary collision ($r_{L} =\sqrt{2}e^2/k_B T$). 
(1) The \emph{unmagnetized regime} ($r_c > \lambda_{\textrm{col}}$) and (2) the \emph{magnetized regime} ($\lambda_D < r_c < \lambda_{D}$) are well described by standard plasma kinetic theories, such as the Landau~\cite{Landau}, Rosenbluth~\cite{Rosenbluth}, Boltzmann~\cite{ferziger1972mathematical}, or Lenard-Balescu~\cite{Lenard,Balescu} equations. 
In these regimes, the Lorentz force may act on the macroscopic distribution function but it does not influence the collision operator. 
(3) The \emph{strongly magnetized} regime ($r_L < r_c < \lambda_D$) has been addressed by extensions of the linear response formalism, such as Rostoker's kinetic equation~\cite{rostoker1960kinetic}, by extensions of the binary collision (Boltzmann) formalism~\cite{Jose_POP_GCO}, and by the Fokker-Planck formalism~\cite{cohen2018fokker,montgomery1974fokker,dubin2014parallel,Ding_Li_2017_Fokker}.
(4) The \emph{extremely magnetized} regime ($r_c < r_L$) has been addressed by the extension of binary collision (Boltzmann) formalism using the generalized Boltzmann kinetic theory~\cite{Jose_POP_GCO} and O'Neil's kinetic theory~\cite{oneil_1983collision}(single component plasmas). 
At strong coupling ($\Gamma > 1$) the parameter space collapses to only two regimes since the Coulomb collision mean free path becomes shorter than both the Debye length and the Landau length. 
Each of the previous approaches addresses a portion of the strong magnetization parameter space at weak coupling conditions ($\Gamma \ll 1$), but do not address strong coupling. 
Here, we show that the combination of the generalized Boltzmann equation from Ref.~\onlinecite{Jose_POP_GCO} and the mean force kinetic equation from Ref.~\onlinecite{Baalrud_PRL} leads to a kinetic theory applicable throughout this parameter space.

In order to test this model, we compute the friction force on a single massive charged particle (ion) moving through a background one-component plasma (electrons). 
We choose the friction force on a single particle as the test transport model for a few reasons: first-principles molecular dynamics simulation data is available with which to benchmark the theory, these data predict novel transport behaviors associated with strong coupling and strong magnetization that a correct kinetic theory should be able to reproduce, friction is a representative transport process that is directly related to macroscopic transport processes such as electrical conduction and diffusion, and it is directly applicable as the stopping rate and range of particles is important in both fusion energy research~\cite{Sigmar_1971_fusion_products} and electron beam cooling experiments~\cite{Men_shikov_2008}. 

The results are found to generally agree well with the MD simulations over the entire range of parameters considered, including weak to strong coupling $\Gamma = 0.1, 1, 10$ and $100$, and weak to strong magnetization $\beta = 0-10$. 
The agreement includes novel effects recently observed in the simulations. 
In particular, the slowing of a test charge due to friction is commonly expected to be described by a single vector component called ``stopping power'' that acts antiparallel to the velocity of the test charge: $F_v = \vc{F} \cdot \hat{\vc{v}}$. 
Previous works~\cite{Jose_POP_GCO,lafleur2019transverse,lafleur2020friction,Derbenev_1978,Men_shikov_2008,parkhomchuk} treating weakly coupled plasmas has shown that in the strongly magnetized and extremely magnetized regimes the friction force has a transverse component in addition to stopping power: $F_{\times} = \vc{F} \cdot (\vc{{\hat{v}}} \times \vc{\hat{n}})$, where $\vc{\hat{n}}$ is the unit vector in direction of the Lorentz force defined as $\vc{\hat{n}} = \vc{\hat{v}} \times \vc{\hat{b}}/\sin \theta$, $\hat{b} = \vc{B}/|\vc{B}|$ is the magnetic field direction and $\theta$ is the angle between the velocity and magnetic field. 
The transverse force is perpendicular to both the velocity and Lorentz force, so it lies in the plane formed by the velocity and magnetic field. The transverse force was initially identified in the strongly magnetized regime for plasmas with temperature anisotropy~\cite{Derbenev_1978,Men_shikov_2008,parkhomchuk} and was later predicted for the Maxwellian plasmas using the linear response theory~\cite{lafleur2019transverse,lafleur2020friction} and confirmed by the molecular dynamics simulations~\cite{David_PRE_2020,Fedotov_MD}. 
Since the transverse force is perpendicular to the velocity, it does not decrease the kinetic energy of the projectile. Thus it was not noticed in the conventional way of obtaining the stopping power by calculating the energy loss of the projectile~\cite{nersisyan2014interactions,Nersisyan_PRE_2000,Nersisyan_PRE_2003,Nersisyan_PRE_2009,cereceda2005stopping}.
The most recent~\cite{David_submitted} MD simulations have extended these studies into the strongly coupled regime. 
These reveal that a third ``gyro friction'' component of the friction force vector arises in the strongly coupled strongly magnetized regime:  $F_n=\vc{F}\cdot \vc{{\hat{n}}}$. 
Here, we show that the kinetic theory is able to capture each of these physical effects. 
The theory is also used to provide an explanation for the physical origin of the gyro friction force, which arises due to asymmetries associated with gyromotion during close collisions.

 Finally, computations of the friction force for all the orientations between the test charge velocity and magnetic field are used to compute an average trajectory of the test charge. 
 It is found that the transverse force strongly influences the trajectory by changing the gyroradius and the range. 
 The gyro friction force is found to slightly alter the gyrofrequency of the test charge resulting in a small phase shift.

The outline of this paper is as follows. In Sec. II, the theoretical formulation and the numerical implementation of the friction force on a projectile is described. Section III compares the results of this calculation with molecular dynamics simulation and discusses the influence of the coupling strength, angle and magnetization on the friction force. In Sec. IV, a qualitative description of the gyro friction force is discussed. Section V describes the trajectories of a projectile moving through a strongly coupled strongly magnetized plasma.

\section{Theory and Evaluation}

The recent generalized Boltzmann kinetic theory showed that particles effectively interact via the potential of mean force during a binary collision and these collisions happen inside a collision volume. The collision volume is a small region of coordinate space where particles interact and its size is determined the range of the potential of the mean force~\cite{baalrud2019mean}. The theory incorporates the gyration of the particles during the collisions by including the Lorentz force in the equations of motion of the colliding particles. The resulting generalized collision operator is

 \begin{eqnarray} \label{gco}
\mathcal{C} =  \chi \int d^3 \vc{v}_2 \int_{S_-} ds  |\vc{u} \cdot  \vc{\hat{s}}|  (f(\vc{v}_1^\prime)f(\vc{v}_2^\prime) - f(\vc{v}_1)f(\vc{v}_2)),
 \end{eqnarray}
where the surface integral is on the surface of the collision volume, $\vc{\hat{s}}$ is the unit normal to the surface and $\vc{u} = \vc{v}_1 - \vc{v}_2$ is the relative velocity of the colliding particles. Here, ($\vc{v}_1$,$\vc{v}_2$) are the precollision velocities, and ($\vc{v}_1^\prime$,$\vc{v}_2^\prime$) are the postcollsion velocities. The postcollision velocities are obtained by solving the equations of motion of interacting particles inside the collision volume. The equations of motion of two charged particles having masses $m_1$ and $m_2$ and charges $e_1$ and $e_2$ interacting in a uniform magnetic field $\vc{B}$ within the collision volume are
\begin{eqnarray}
m_{1} \frac{d\vc{v}_{1}}{dt} &=& -\nabla_{\vc{r}_1}\phi(r)+ e_1\Big(\frac{\vc{v}_{1}}{c}\times\vc{B}\Big) \label{first} \\
m_{2} \frac{d\vc{v}_{2}}{dt} &=& -\nabla_{\vc{r}_2}\phi(r) +e_2\Big(\frac{\vc{v}_{2}}{c}\times\vc{B}\Big) . \label{second}
\end{eqnarray}
Here, $\phi(r)$ is the potential of mean force. 

 When the plasma is strongly coupled, spatial correlations between the particles become significant. The generalized kinetic theory accounts for this many-body effect in the collision by using the potential of mean force, which is the potential obtained by fixing the positions of two particles at a distance $r$ apart and averaging over the positions of the remaining particles at equilibrium. The potential of mean force asymptotes to the Debye-H\"{u}ckel potential in the weakly coupled limit and was used to compute the friction force in Ref.~\onlinecite{Jose_POP_GCO}. In the strongly coupled regime, the potential of mean force for a one-component plasma can be accurately modeled using the hypernetted-chain approximation (HNC)~\cite{hansen2013theory}.
\begin{subequations}
\begin{eqnarray}
g(r) &=& \exp[v(r)+h(r)-c(r)]   , \label{phi1} \\
\hat{h}(k) &=& \hat{c}(k)[1+n\hat{h}(k)],\label{phi2} 
\end{eqnarray}
\end{subequations}
where $h(r) = g(r)-1$, $\hat{h}(k)$ is the Fourier transform of $h(r)$, $v(r) = e^2/r$ is the Coulomb potential, $c(r)$ is the direct correlation function and $\hat{c}(k)$ is its Fourier transform. The potential of mean force is obtained from the pair distribution function via: $\phi(r) = -k_B T\ln(g(r))$. 

The use of the potential of mean force to model the binary interactions introduces aspects of many-body effects such as screening and correlations that become essential at strong coupling. The mean force kinetic theory also accounts for the Coulomb hole surrounding the particles interacting via Coulomb force~\cite{Baalrud_enskog}. This excluded volume leads to a reduced volume of space that the particles can occupy resulting in an increased collision frequency for strongly coupled plasma. This leads to a frequency enhancement factor $\chi = g(r = \sigma)$, where $\sigma$ is the Coulomb hole radius, in the collision operator. This is obtained from the modified version of the Enskog's theory of hard spheres for plasmas~\cite{Baalrud_enskog}. The $\chi$ factor for the coupling strengths studied here $\Gamma = 1$, $10$ and $100 $ are $1.36$, $1.45$ and $1.65$ respectively.

To illustrate the utility of the generalized collision operator, we compute the friction force on a massive projectile moving through a strongly coupled strongly magnetized one-component plasma. The friction force is $\vc{F}=\vc{R}^{12}/n_1$, where $\vc{R}^{12}$ is the friction force density obtained by taking the momentum moment of the collision operator ($\int d^3 \vc{v}_1 \mathcal{C}$) and $n_1$ is the density of the projectile. Using the generalized collision operator (Eq. (\ref{gco})),

 \begin{eqnarray}
\vc{R}^{12} =  \frac{\chi n_1 n_2 m_1}{\pi^{3/2}v_T^3}\int d^3 \vc{v}_2 \int_{S_-}  ds \, |\vc{u} \cdot  \vc{\hat{s}}|   \nonumber \\
(\vc{v}_1^\prime-\vc{v}_0)\exp\bigg(\frac{-v_2^2}{v_T^2}\bigg). \label{Friction}
\end{eqnarray}
Here, the distribution function of the projectile is taken as the Dirac delta function: $f_1 = n_1 \delta^3(\vc{v}_1-\vc{v}_0)$, where $\vc{v}_0$ is the initial velocity of the projectile and the velocity distribution function of the background plasma is taken as Maxwellian:  $f_2 = \frac{n_2}{\pi^{3/2}v_T^3}\exp(-v_2^2/v_T^2)$, where $n_2$ is the plasma density and $v_T=\sqrt{2 k_B T/m_2}$ is the thermal velocity. We consider a massive projectile with mass ratio $m_r = m_1/m_2 = 1000$ and a charge that is same as that of the background plasma particles ($e$). During a binary collision between the projectile and the background plasma particle, they both are acted upon by the Lorentz force from the external magnetic field in addition to the interaction via potential of mean force. However, since the projectile is very massive the effect of the Lorentz force on its motion is negligible compared to that of the background particles and thus this term is accurately neglected from the equations of motion. The equations of motion~(Eqs. (\ref{first}) and (\ref{second})) in relative and center of mass velocities then reduce to
 \begin{eqnarray}
&&(m_1+m_2)\frac{d\vc{V}}{dt} = e \Big(\frac{\vc{V}}{c} \times \vc{B}\Big)-\frac{e \, m_{12}}{m_2} \Big(\frac{\vc{u}}{c} \times \vc{B}\Big)\label{main1} \\
&&m_{12}\frac{d\vc{u}}{dt} = -\nabla \phi(r)+ \frac{e \, m_{12}^2}{m_2^2} \Big(\frac{\vc{u}}{c} \times \vc{B}\Big)  \notag \\ 
&& \, \,\, \,\, \,\, \, \, \, \, \, \, \, \, \, \, \, \, \, \, \, \, \, \, \, \,  -\frac{e \, m_{12}}{m_2}  \Big(\frac{\vc{V}}{c} \times \vc{B}\Big) .\label{main2}
\end{eqnarray}
where $\vc{V} = m_1 \vc{v}_1+m_2 \vc{v}_2/(m_1+m_2)$ is the center of mass velocity and $m_{12} = m_1m_2/(m_1+m_2)$ is the reduced mass.

\begin{figure*} [!htb] 
\centerline{\includegraphics[width = 7.5in]{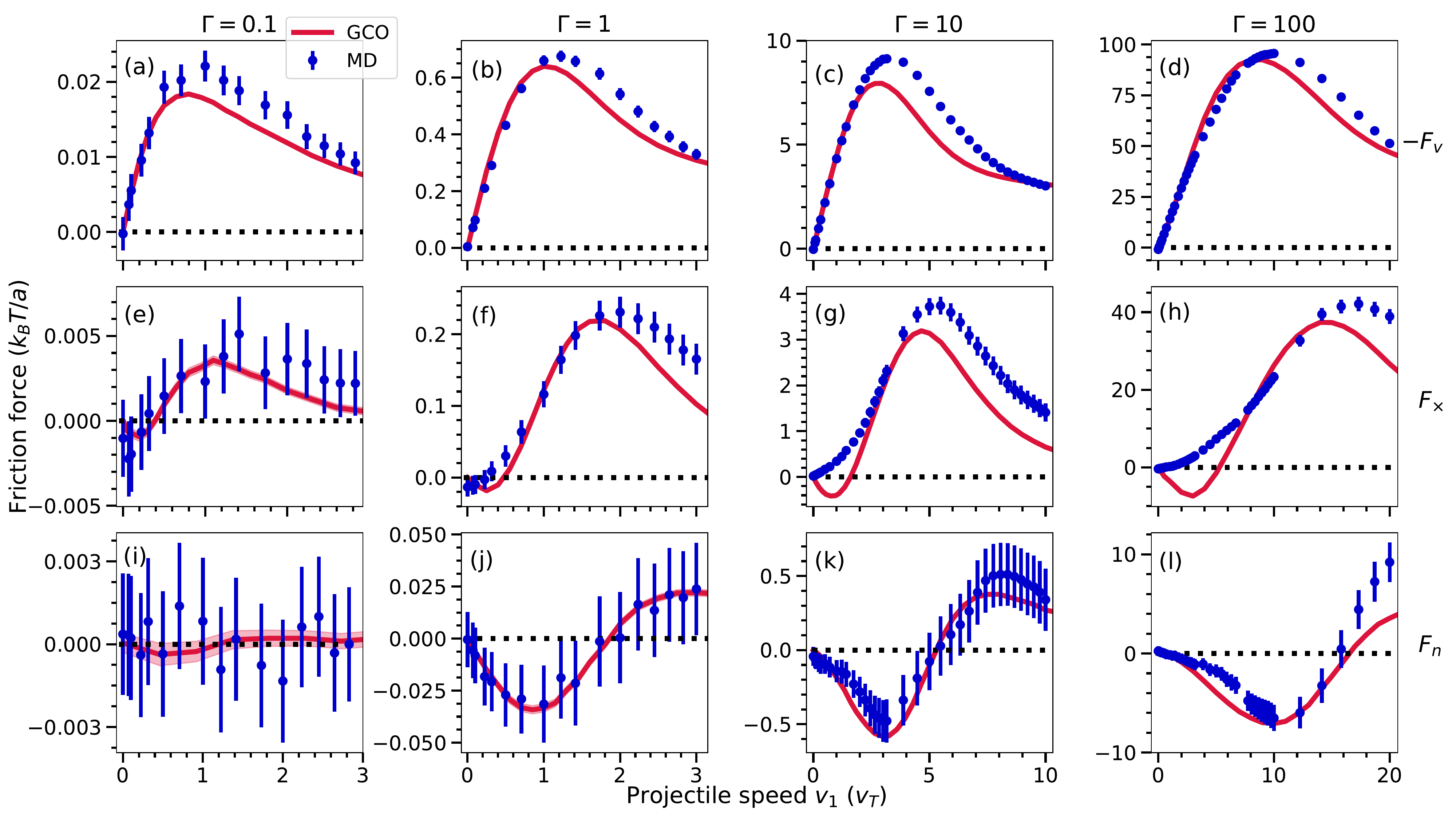}}
\caption {Friction force components at $\beta =10$ and $\theta = 22.5 \degree $ for different coupling strengths $\Gamma = 0.1 $ [a, e, and i], $\Gamma = 1 $ [b, f, and j], $\Gamma = 10 $ [c, g, and k]and $\Gamma = 100 $ [d, h, and l]. The generalized collision operator (GCO) results are shown as red solid lines and the molecular dynamics (MD) results from Ref.~\onlinecite{David_submitted} as data points. }
  \label{fig:Gammas}
\end{figure*}

 \begin{figure} [!htb] 
\centerline{\includegraphics[width = 1.0\columnwidth]{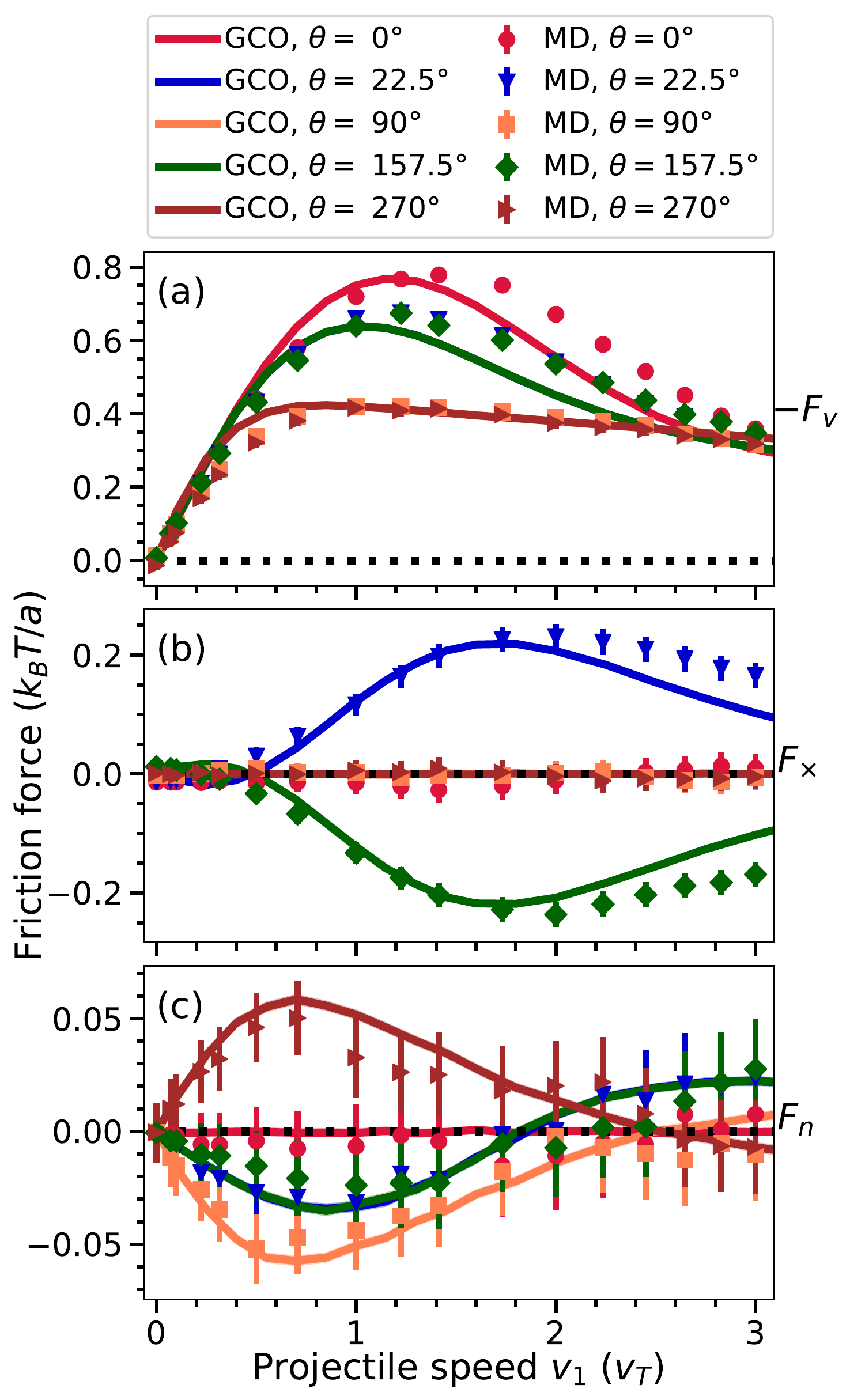}}
\caption {Comparison of GCO (lines) and MD (data points) predictions for the friction force components at $\Gamma =1$ and $\beta=10$ for different orientations of projectile and magnetic field $\theta = 0 \degree $, $ 22.5 \degree $, $90 \degree $, $ 157.5 \degree $ and $ 270 \degree $.}
  \label{fig:Angles}
\end{figure}

 \begin{figure} [!htb] 
\centerline{\includegraphics[width = 1.0\columnwidth]{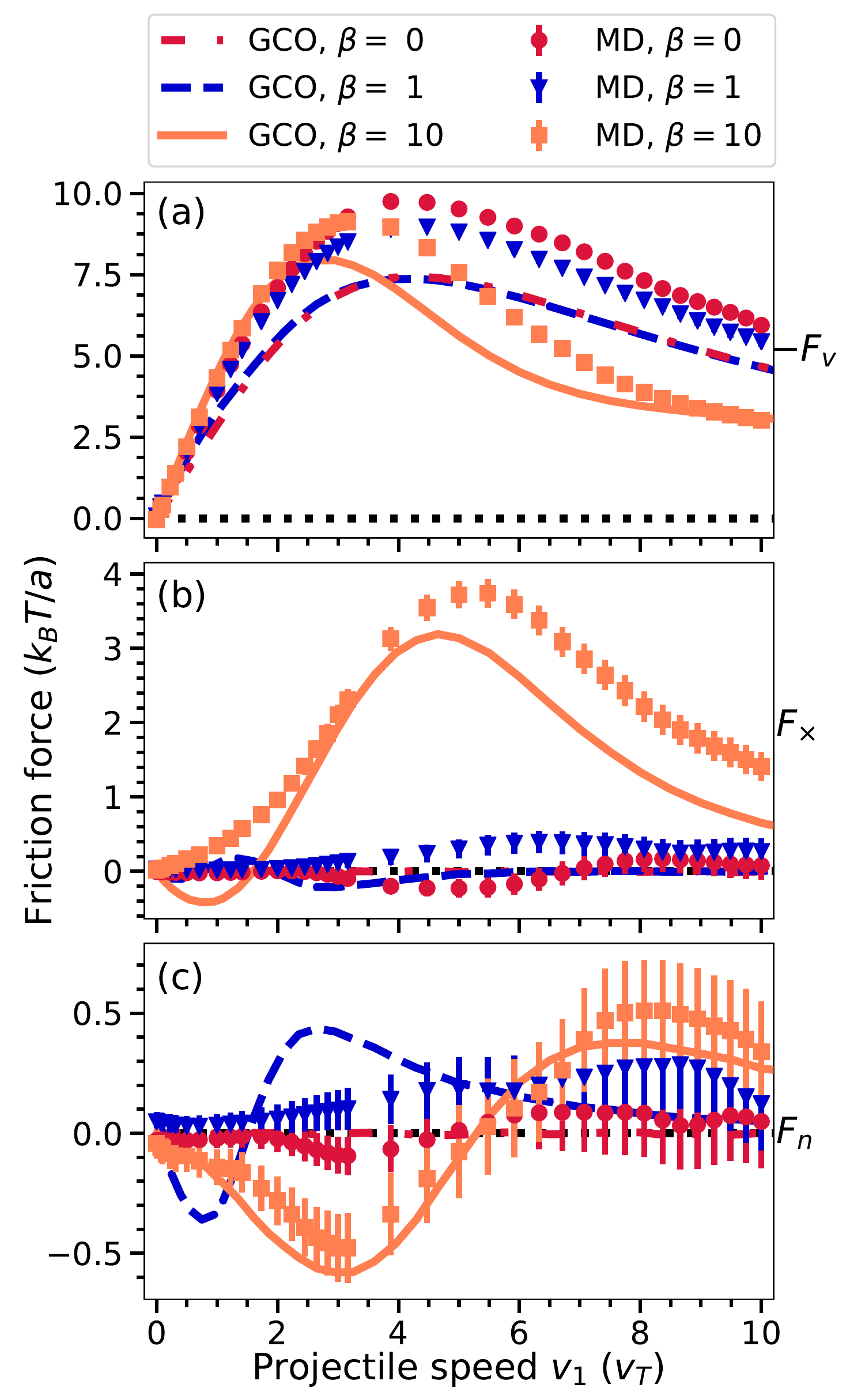}}
\caption {Comparison of GCO (lines) and MD (data points) predictions for the friction force components at $\Gamma =10$ and $\theta =22.5 \degree $ for different magnetization strengths $\beta = 0$, $ 1$ and $10$.}
  \label{fig:Betas_MD}
\end{figure}

The potential of mean force ($\phi(r)$) was computed numerically by solving Eqs.~(\ref{phi1})~and~(\ref{phi2}). The result was interpolated using the Cubic spline method~\cite{press2007numerical} for the trajectory calculations. Numerical evaluation of the friction force integral (Eq.~(\ref{Friction})) was the same as that described in Ref.~\onlinecite{Jose_POP_GCO} except that the numerically computed potential of mean force was used instead of the Debye-H\"{u}ckel potential, and the variables used to solve the equations of motion inside the collision volume were scaled using the Wigner-Seitz radius~($a$) instead of the Debye-length~($\lambda_D$) to account for the change in the scale of the collision volume at strong coupling.


\begin{figure*} [!htb] 
\centerline{\includegraphics[width = 7.5in]{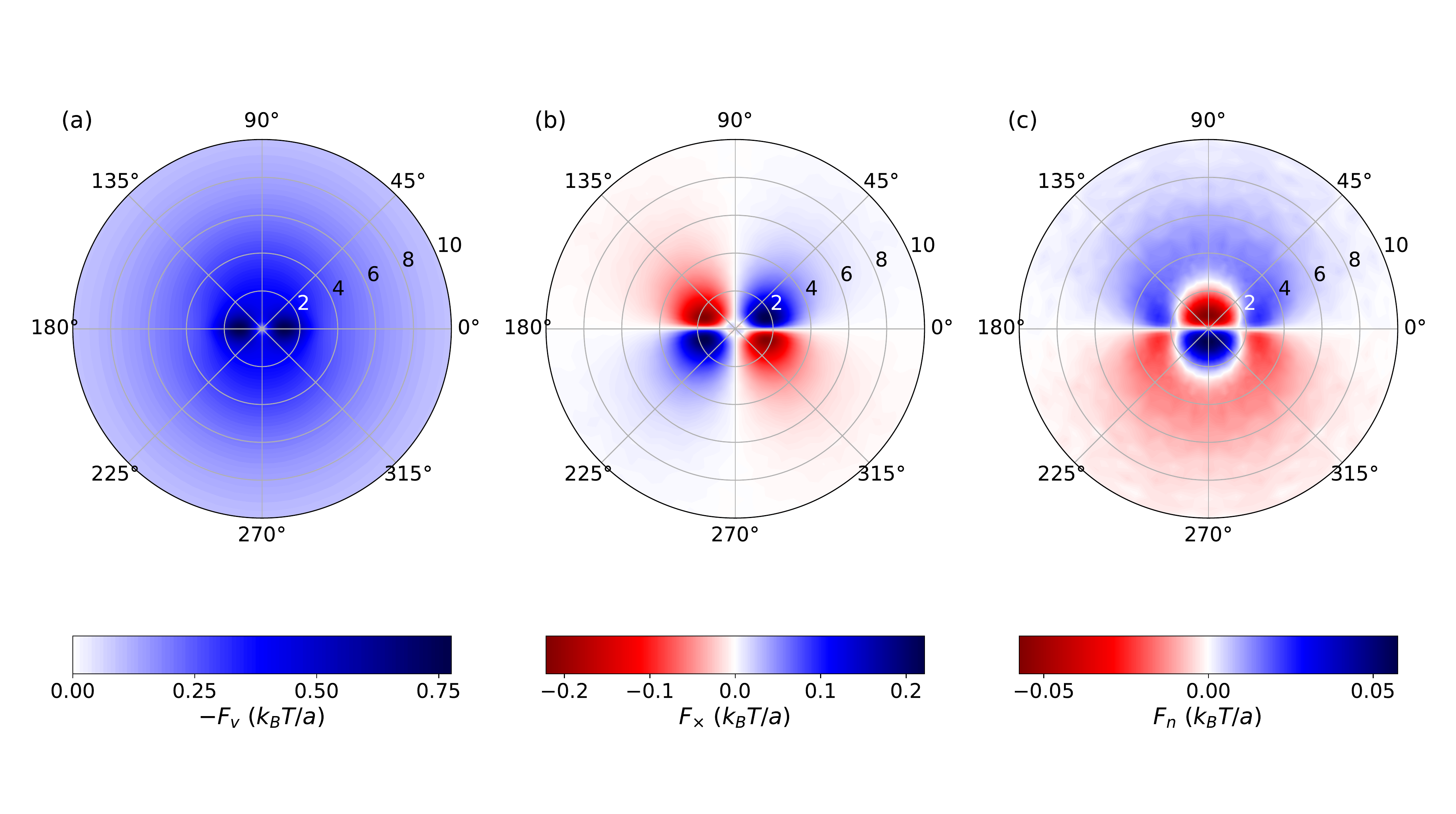}}
\caption {Polar plots of the friction force components [$-F_v$ in panel (a), $F_\times $ in panel (b) and $F_n$ in panel (c)] at $\Gamma=1 $ and $\beta = 10$. The radial axis is the speed of the projectile and the angle is the phase angle that the projectile's velocity makes with the direction of the magnetic field.}
  \label{fig:Polar_Gamma_1}
\end{figure*}

\begin{figure*} [!htb] 
\centerline{\includegraphics[width = 7.5in]{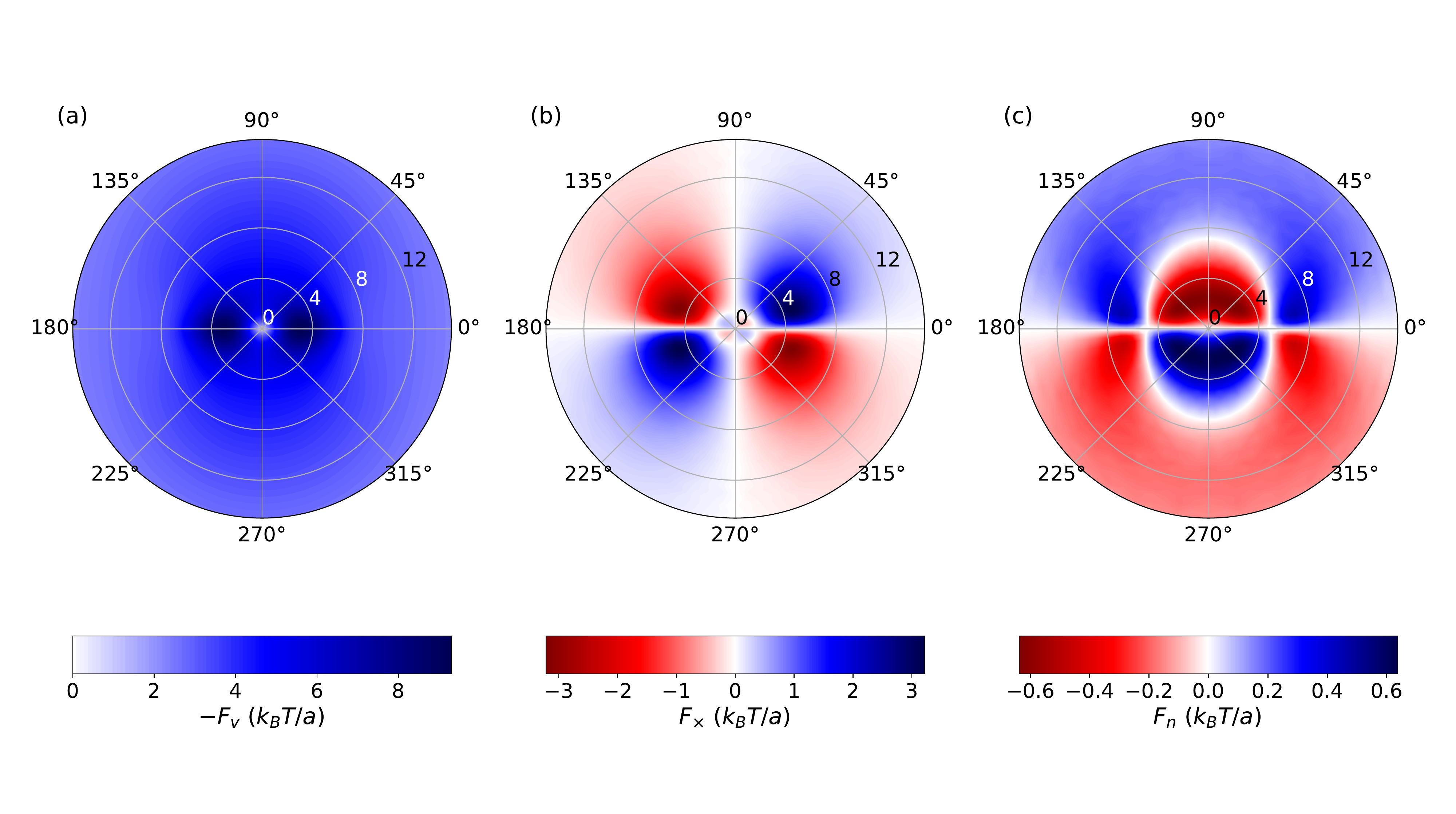}}
\caption {Polar plots of the friction force components [$-F_v$ in panel (a), $F_\times $ in panel (b) and $F_n$ in panel (c)] at $\Gamma=10 $ and $\beta = 10$. The radial axis is the speed of the projectile and the angle is the phase angle that the projectile's velocity makes with the direction of the magnetic field.}
  \label{fig:Polar_Gamma_10}
\end{figure*}

\begin{figure*} [!htb] 
\centerline{\includegraphics[width = 7.5in]{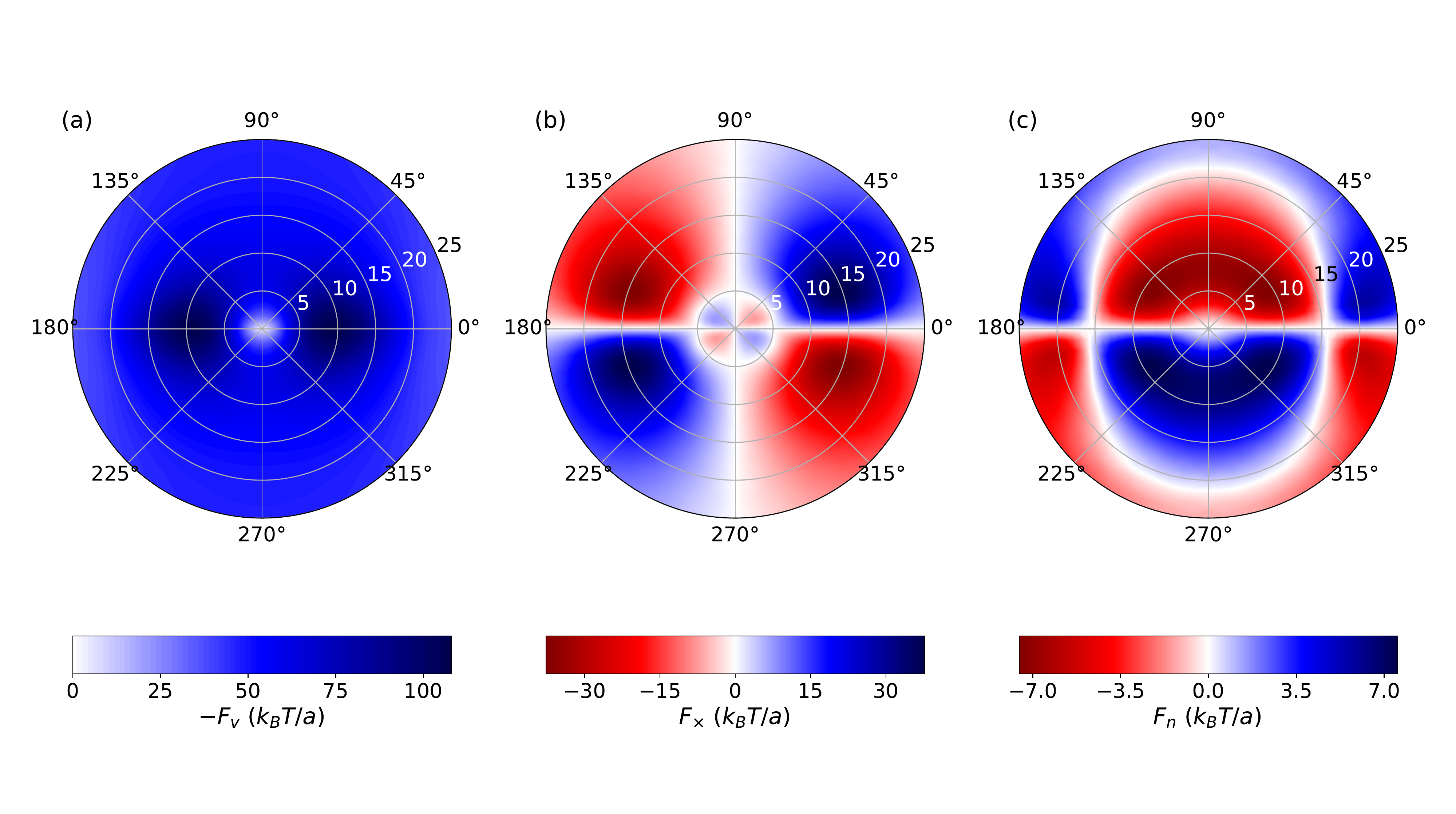}}
\caption {Polar plots of the friction force components [$-F_v$ in panel (a), $F_\times $ in panel (b) and $F_n$ in panel (c)] at $\Gamma=100 $ and $\beta = 10$. The radial axis is the speed of the projectile and the angle is the phase angle that the projectile's velocity makes with the direction of the magnetic field.}
  \label{fig:Polar_Gamma_100}
\end{figure*}

 \begin{figure} [!htb] 
\centerline{\includegraphics[width = 1.0\columnwidth]{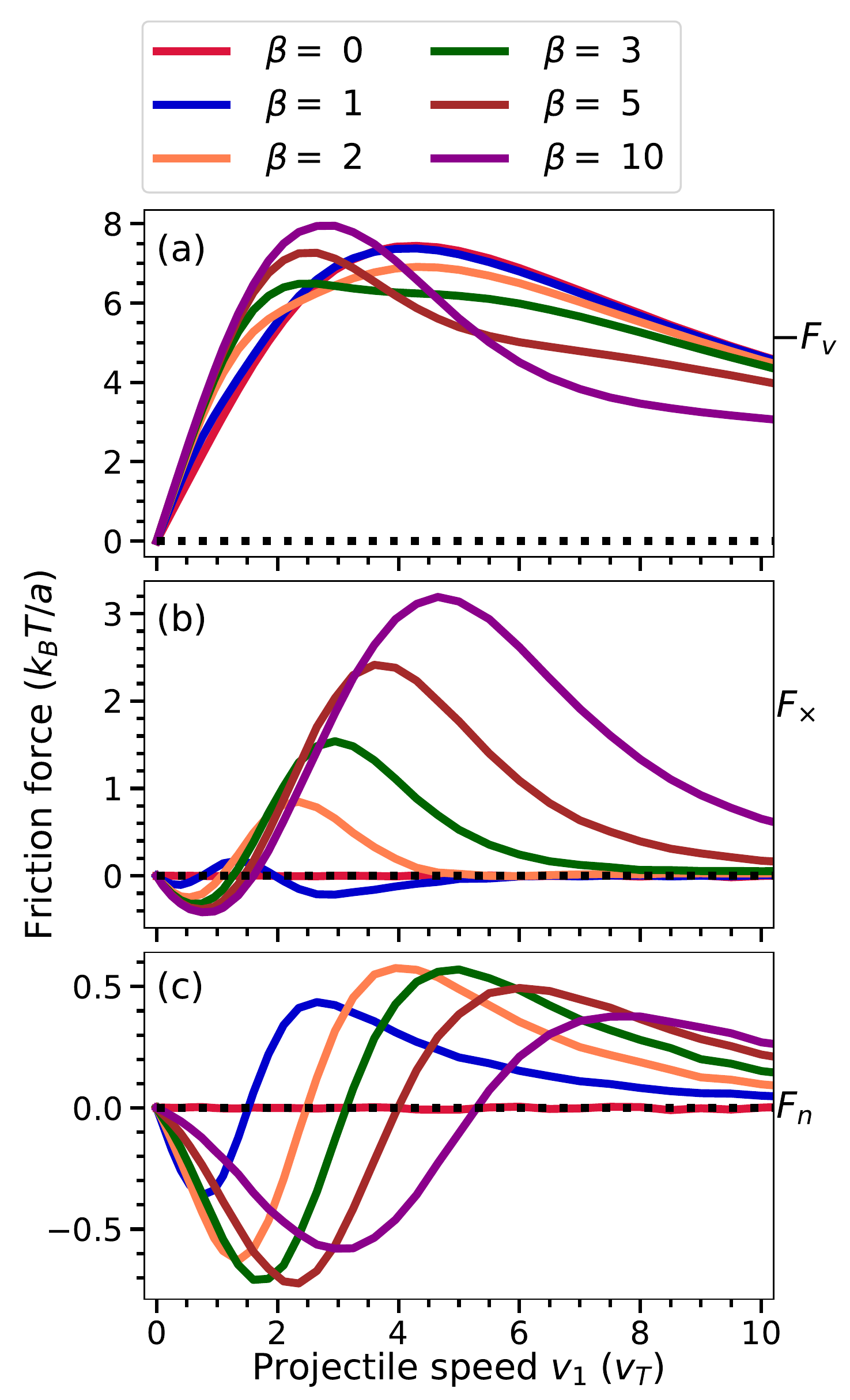}}
\caption {Friction force components at $\Gamma =10$ and $\theta =22.5 \degree $ for different magnetization strengths $\beta = 0$, $1$, $2$, $3$, $5$ and $10$.}
  \label{fig:betas_GCO}
\end{figure}
\section{Results}

\subsection{Comparison with Molecular Dynamics Simulations}

In order to test the results computed from the generalized collision operator (Eq.~(\ref{Friction})), we compare with results from recent molecular dynamics simulations~\citep{David_submitted}. Molecular dynamics simulations provide a rigorous benchmark because they directly solve the first-principles equations of motion for all interacting particles. The results are compared across different coupling strengths, magnetization strengths and orientation of the projectile's velocity with respect to the direction of magnetic field.

Figure \ref{fig:Gammas} compares the friction force curves by fixing the magnetization strength ($\beta =10$) and orientation of projectile's velocity with respect to the direction of magnetic field ($\theta = 22.5\degree$) and varying the coupling strength ($\Gamma = 0.1,1,10$ and $100$). The model predictions are generally in good agreement with the MD results across this entire range of coupling strengths. At $\Gamma =100$, the plasma is liquid-like and the agreement between theory and MD is quite remarkable. In these plots, $\Gamma = 0.1 $ are in the strongly magnetized transport regime and the rest are in the extremely magnetized transport regime as defined in the introduction. This shows the versatility of the GCO to be applicable across a wide range of both Coulomb coupling and magnetization strength regimes. Previous theoretical approaches~\cite{lafleur2019transverse,lafleur2020friction} based on linear response theory is limited to the strongly magnetized transport regime due to the inadequacy of the theory to account for strong non-linear interactions that arise in either in the extremely magnetized regime, or the strongly coupled regime. 

 On comparing the stopping power curves, good qualitative agreement is observed. Features like an increase in magnitude of the force, shift in position of the Bragg peak and broadening of the curves with increasing coupling strength are observed. A good quantitative agreement between theory and MD is observed for small speeds of the projectile, but some quantitative differences emerge at high speeds.  This was also observed in unmagnetized plasmas~\cite{bernstein2019effects}. The larger discrepancy at high speeds may be due to the absence of dynamic screening (velocity-dependent screening) in the potential of mean force used for modeling the binary interactions.

The transverse force curves obtained by the theory also capture the qualitative trends predicted by the MD. Features like an increase in the magnitude of the force and broadening of the curves with increasing coupling strength are observed in both. However, some quantitative differences are observed. The most significant difference is that for coupling strengths $\Gamma =10$ and $\Gamma=100$, the theory predicts a change in sign of the transverse force at low speeds, which is not observed in the MD data. This change in sign is a prominent characteristic of the transverse force and is observed when $\Gamma =0.1$ and $\Gamma=1$ by both the theory and MD. The GCO calculations predict that the change in sign happens at higher speeds with increasing coupling strength. The cause of this disagreement remains uncertain at this time. Strong magnetization causes particles to move tightly along the field lines and to re-collide multiple times, resulting in increased inter-particle correlations in length and time~\cite{Keith_POP_2021}. These increased correlations increase the time it takes for the plasma to reach hydrodynamic behavior. One possible reason for the discrepancy might be that the strong correlations cause a disconnect between the concept of an ``instantaneous'' friction force with what is computed over a few plasma period interval in the MD simulations. Such correlations may also violate the molecular chaos approximation used in the generalized collision operator~\cite{Dubin_recollision} and might be another reason for the discrepancy. 


A significant result of this work is the prediction of the gyro friction force component in the direction of the Lorentz force ($F_n$). Good agreement is found between the theory and the MD across coupling strengths. This component is found to be negligible in the weakly coupled regime and its strength increases with the coupling strength. This prediction shows the ability of the generalized collision operator to accurately capture novel physics arising from the combination of strong magnetization and strong coupling.

The friction force is found to not only depend on the speed of the projectile but also on the orientation of the projectile's velocity with the direction of the magnetic field. Figure \ref{fig:Angles} compares the friction force curves for different orientation of the projectiles velocity with the direction of the magnetic field ($\theta = 0\degree$, $22.5 \degree$, $90 \degree$, $157.5\degree$ and $270 \degree$) for the coupling strength $\Gamma =1$ and the magnetization strength $\beta =10$. The good agreement with MD shows that the theory can accurately capture the dependence of the friction force components on the orientation of the projectile's velocity. 


At strong coupling, the coupling-magnetization parameter space identifying transport regimes is predicted to collapse to two regions  - unmagnetized and extremely magnetized. Fixing the coupling strength $\Gamma$ and changing the magnetization strength $\beta$ can span these two transport regimes. Figure \ref{fig:Betas_MD} compares the friction force components obtained using MD simulations and theory for magnetization strengths $\beta = 0, 1$ and $10$. Here the coupling strength is $\Gamma=10$ and angle $\theta = 22.5 \degree$. Theory captures the overall trends observed in the MD results. Both predict that strong magnetization causes a shift of the position of the Bragg peak to lower speed and decrease the stopping power at high speeds. The increase in relative magnitude of the transverse force with the increase in magnetization strength is captured by both the theory and MD. However, some quantitative features like the position of the peaks and sign reversal differ. For instance, the theory predicts sign reversal of the transverse force (once for $\beta=10$ and twice for $\beta=1$), which is absent in the MD results.  Good agreement between the theory and MD is seen in the gyro friction force curve for $\beta =10$, and both are consistent with zero at $\beta = 0$. However, at the transition magnetization strength ($\beta =1$), the agreement is poor across the projectile speeds. 

\subsection{Coupling Strength and Angle}\label{polar}
 
 The friction force exhibits strong dependence on the orientation of the velocity with the direction of the magnetic field. For this reason, an entire polar plot that includes both speed and angle is required to compute the average trajectory of a test charge in a strongly magnetized plasma. Because so many data points are required to solve for this 2D parameter space, it is impractical to obtain this information from MD simulations due to their high computational cost. However, the much lower computational expense of the GCO calculations makes this possible. Polar plots also reveal the symmetry properties of the different friction force components.

  
 Figures \ref{fig:Polar_Gamma_1}, \ref{fig:Polar_Gamma_10}, and \ref{fig:Polar_Gamma_100} show the polar plots of each friction force component for $\beta= 10$ and coupling strengths $\Gamma = 1$, $10$ and $100$. On comparing  these figures, quantitative changes are observed. The magnitude of the friction increases with increasing coupling strength.  
Panel (a) of figures \ref{fig:Polar_Gamma_1}, \ref{fig:Polar_Gamma_10}, and \ref{fig:Polar_Gamma_100} show that the stopping power curve broadens and the Bragg peak shifts to higher speed with increasing coupling strength. This trend is similar to that observed in the unmagnetized plasma~\cite{bernstein2019effects,baalrud2016effective}. The phase angle of the stopping power follows the symmetry: $F_v (\theta) = F_v (\pi-\theta) = F_v (\pi+\theta) = F_v (2 \pi-\theta)$ for $0  \leq \theta \leq \pi /2$. This is the same symmetry observed in the weakly coupled transport regime~\cite{lafleur2019transverse}. When $\theta$ increases from $ 0 \degree$, the magnitude of the Bragg peak monotonically decreases and reaches a minimum value at $90 \degree$. The position of the Bragg peak is observed to shift to lower speed with increasing $\theta$ in the first quadrant. Since stopping power is anti-parallel to the velocity of the projectile, these results imply that the energy deposition of the projectile has a significant dependence on the orientation of the velocity vector.

Panel (b) of figures \ref{fig:Polar_Gamma_1}, \ref{fig:Polar_Gamma_10}, and \ref{fig:Polar_Gamma_100} show the transverse component of the friction. Similar to the stopping power, both the positive and negative peaks shift to higher speeds and the transverse force curve widens as the coupling strength increases. The prominent sign reversal signature of the transverse force is seen across coupling strengths. The phase angle of the transverse force follows the symmetry: $F_{\times} (\theta) = -F_{\times} (\pi-\theta) = F_{\times} (\pi+\theta) =- F_{\times} (2 \pi-\theta)$ for $0  \leq \theta \leq \pi /2$. This is the same symmetry observed at weakly coupling~\cite{lafleur2019transverse}. Similar to the previous findings, the transverse force is zero when the projectile's motion is parallel or perpendicular to the magnetic field. For a phase angle less than $90 \degree$, a positive transverse force increases the gyroradius and a negative transverse force decreases the gyroradius. Thus the transverse force increases the gyroradius of the fast projectile by redirecting the kinetic energy from the parallel direction to the perpendicular direction and decreases the gyroradius of the slower projectile by redirecting the kinetic energy from the perpendicular direction to the parallel direction.

Panel (c) of figures \ref{fig:Polar_Gamma_1}, \ref{fig:Polar_Gamma_10}, and \ref{fig:Polar_Gamma_100} show the gyro friction component of the friction force. This component is absent in linear response theory calculations~\cite{lafleur2019transverse,lafleur2020friction}, which apply in the weakly coupled regime. The linear response theory inherently assumes that the interactions between the colliding particles are weak and avoids the strong non-linear interactions characterizing close collisions~\cite{nicholson1983introduction}. However, the binary collision theories are capable of capturing the physics associated with the strong close collisions. This suggests that the $F_n$ component of the friction force is due to close collisions. 

The magnitude of the gyro friction component is found to be smaller than the other two components for all coupling strengths studied. Similar to the transverse force, the sign of the gyro friction force has a dependence on the speed of the projectile. The critical speed at which the transition occurs is found to depend on the coupling strength. Similar to the other two friction components, the force curve broadens and both the positive and negative peaks shift to higher speeds with the increase in coupling strength. The phase angle of the gyro friction follows the symmetry: $F_{n} (\theta) = F_{n} (\pi-\theta) = -F_{n} (\pi+\theta) =- F_{n} (2 \pi-\theta)$ for $0  \leq \theta \leq \pi /2$. The gyro friction is zero when the projectile moves parallel or anti-parallel to the magnetic field ($\theta = 0\degree$ and $\theta = 180 \degree$). Its magnitude is maximum when the projectile's motion is perpendicular to the magnetic field. For a phase angle less than $90 \degree $, a positive sign of the gyro friction corresponds to an increase in the gyrofrequency of the projectile, whereas a negative sign corresponds to a decrease in the gyrofrequency. Thus, gyro friction increases the gyrofrequency of a fast projectile and decreases that of a slow projectile.

\subsection{Magnetization}
 In order to study how the friction force components change across transport regimes, we compute the friction force curves for a fixed coupling strength ($\Gamma = 10$) and orientation of the projectile's velocity ($\theta = 22.5 \degree$) and change the magnetization strength ($\beta = 0$ to $10$). The boundary between the unmagnetized and extremely magnetized transport regimes for $\Gamma =10$ is at $\beta \approx 0.25$. 
 
 Figure \ref{fig:betas_GCO} shows the three components of the friction force curves for $\beta = 0,1,2,3,5$ and $10$. The Bragg peak of the stopping power curve shifts to the lower speeds and the magnitude of the Bragg peak increases on moving from the unmagnetized to the extremely magnetized regime. Careful examination shows that this transition involves 2 stages. A new peak develops at low speed and increases in magnitude with increasing magnetization strength. Simultaneously, the stopping power at high speed decreases. Thus magnetization increases the stopping power of slow projectiles and decreases the stopping power of fast projectiles. This is similar to the weak coupling limit~\cite{lafleur2019transverse}. 

The transverse force is present only when the plasma is in the extremely magnetized transport regime and its magnitude increases with increasing magnetization strength. For this orientation of the projectile ($\theta = 22.5 \degree$) the positive sign of the transverse force corresponds to a force that acts to increase the gyroradius of the projectile, and the negative sign corresponds to a force that decreases its gyroradius. The curve for $\beta = 1$ predicts two negative dips of very small magnitudes. This prediction is unique to the regime of strong coupling and a transitional magnetic field strength ($\beta \approx 1$). The effect may be associated with the non-monotonic nature of the potential of mean force at strong coupling. The second negative dip disappears with increasing magnetization strength.

The gyro friction is absent in the unmagnetized regime ($\beta = 0$). For this angle ($\theta = 22.5$), a negative sign of the gyro friction corresponds to a force that acts to decrease the gyrofrequncy of the projectile, whereas a positive sign corresponds to a force that acts to increase its gyrofrequency. The increase in magnetization strength from $\beta =1$ to $\beta =10$, increases and then decreases the magnitude of the peak of the gyro friction. This is in contrast to $F_{\times}$ where the magnitude of the peak monotonically increases. However the change in the critical speed at which the gyro friction changes its direction has a strong dependence on the magnetization strength. The critical speed moves to larger speeds with increasing magnetization strength. This can also be contrast with the transverse force, where the sign change is nearly independent of $\beta$.

\begin{figure*} [!htb] 
\centerline{\includegraphics[width = 7.5in]{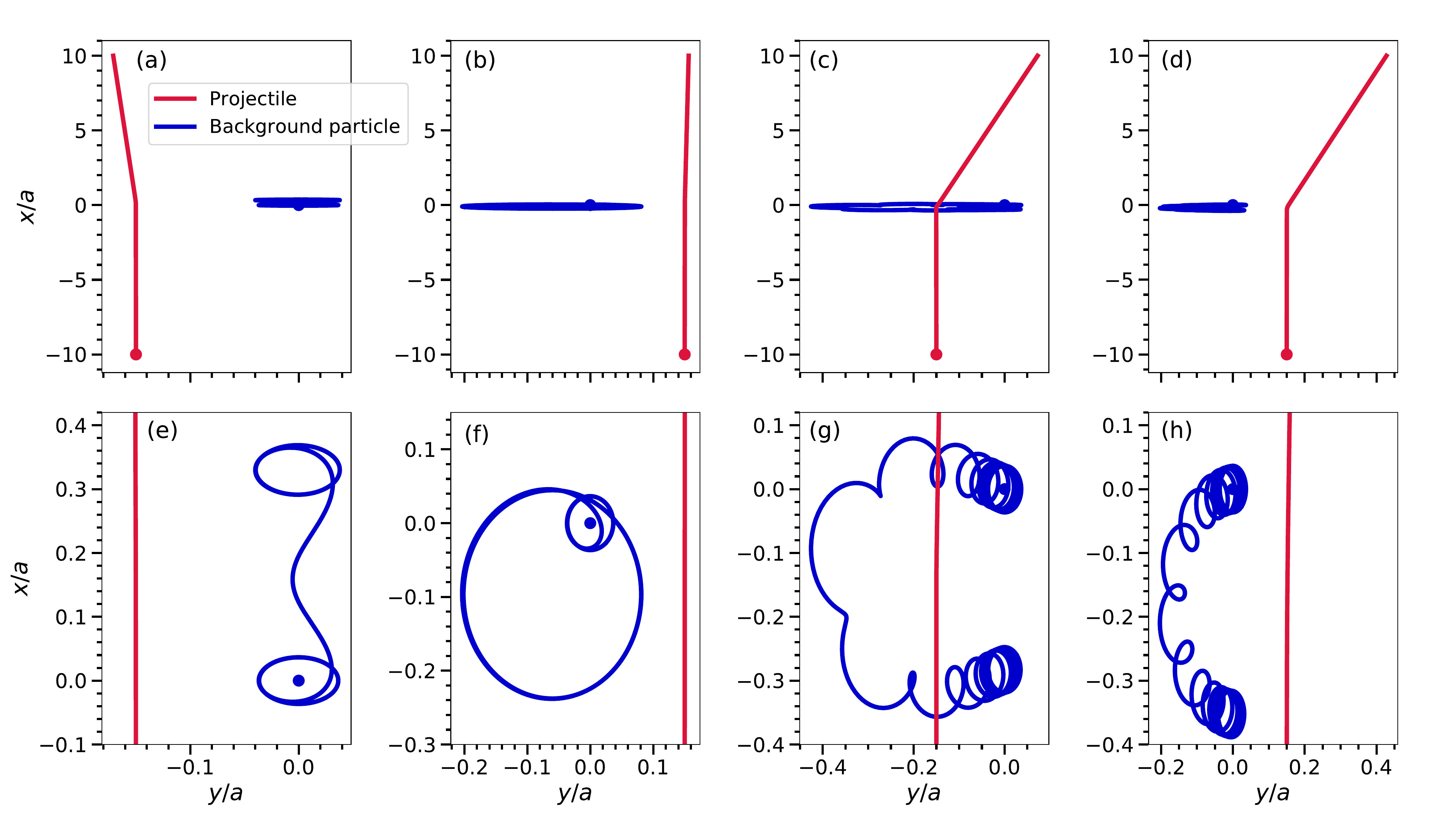}}
\caption { Trajectories of the projectile (red) and the background plasma particle (blue) during a Coulomb collision. The magnetic field is in the $\hat{z}$ direction and the parallel speed of both the projectile and the background particle are taken as zero. Panels (a), (b), (e) and (f) show the case of a fast projectile ($v_1 = 6 v_T$) and panels (c), (d), (g) and (h) are for a slow projectile ($v_1 = 0.5 v_T$). The bottom panels show the zoomed in view of the top panels. The initial speed of the background particle is taken as ($1 v_T$, $1 v_T$, $0$) and the initial guiding center position (blue dot) of the background particle is at (0, 0, 0). The initial position of the projectiles (red dot) are ($-10a$, $-0.15a$, 0) [(a), (e), (c), (g)] and ($-10a$, $0.15a$, 0) [(b), (f), (d), (h)]. The trajectories are for $\Gamma =10 $ and $\beta = 10$ and interactions are modeled using the Debye-H\"{u}ckel potential. }
  \label{fig:Discussion}
\end{figure*}

\section{Discussion}
 A qualitative description of the physical origin of the stopping power and the transverse force from the binary collision perspective was provided in Ref.~\onlinecite{Jose_POP_GCO}. Although that reference concentrated on the weakly coupled regime, the qualitative explanations carry over to the strongly coupled regime. Here, we give a qualitative description of the physical origin of the gyro friction.  
 
 The gyro friction is only observed in the strongly coupled strongly magnetized regime. When the plasma is strongly coupled, the screening distance reduces and the close collisions become more relevant. The gyro friction is also absent in the linear response calculations, which ignores close collisions. These suggest that the gyro friction is associated with close collisions. The binary collision theory obtains the friction force by summing the change in momentum of the projectile from all possible binary scattering events. In order to have a clearer qualitative description, consider the interaction of the test charge and a background particle in 2D, corresponding to the situation of the particles having no parallel velocity. 
The origin of the gyro friction force can be understood by considering violations of symmetry for interactions in which the gyrocenter of the background particle starts at a fixed distance either to the right or left of the projectile. 
Since the net force is the result of the sum of the collision from both sides, any asymmetry will lead to a net force in the $y$ direction. 
The dominant symmetry breaking process is observed to depend on the speed of the projectile compared to the background particle. 
At sufficiently high speed it is associated with an asymmetry in the relative velocity between the projectile and background particle when they are closest together (most strongly interacting). 
At sufficiently low speed it is is associated with an $\vc{E} \times \vc{B}$ drifting motion of the background particle. 
Each process is found to lead to a different sign of the gyro friction and the sign change is associated with the transition from one process dominating over the other. 


These effects are illustrated in the example trajectories shown in Fig.~\ref{fig:Discussion}. Here, the guiding center of the background particle is (0, 0, 0), and the projectile starts equidistantly either to left or to right of the guiding center. The magnetic field is in the  $+\hat{z}$ direction and the charge of the projectile and the background particle is positive. This makes the background particles gyrate in the counterclockwise direction. The initial velocity of the projectile is in the $\hat{x}$ direction, thus  $\hat{n}=-\hat{y}$.

First, consider the interaction of a fast projectile with a thermal background particle; as shown in panels (a), (b), (e) and (f) of Fig.~\ref{fig:Discussion}. 
When the projectile starts to the left of the background particle (panels (a) and (e)), the collision deflects the projectile to the left, resulting in a change in momentum in the $-\hat{y}$ direction. In contrast, when the projectile starts to right of the background particle (panels (b) and (f)), the collision deflects the projectile to the right, resulting the change in momentum in the $+\hat{y}$ direction. 
However, the momentum transfer is greater when the projectile approaches from the left than from the right. 
This asymmetry causes a net force in the $-\hat{y}$ direction, which is the $+\hat{n}$ direction, leading to a positive sign of the gyro friction at high speed. 
The reason for the asymmetry is that the relative velocity between the projectile and background particle is higher when the projectile approaches from the left because the background particle gyrates counterclockwise. 
Within the region of closest approach, where the interaction is strongest, the interaction is head-on when the approach is from the left. 
In contrast, the $\hat{x}$ component of the velocity of both particles is positive when the projectile approaches from the right. 
The higher relative velocity in the head-on collision causes a greater momentum transfer. 
Any $\vc{E} \times \vc{B}$ motion is negligible in this case because the collision time is too short for the background particle to exhibit an $\vc{E} \times \vc{B}$ drift when the projectile speed is large.  




In contrast, panels  (c), (d), (g) and (h) of Fig.~\ref{fig:Discussion} show the collision of a slow projectile with a thermal particle of the background plasma. 
At these conditions, the projectile is observed to deflect to the right ($+\hat{y}$ direction) whether it starts to the right or left of the  background particle.
The reason for this qualitative difference is that when the projectile approaches slowly, the background particle has enough time to $\vc{E} \times \vc{B}$ drift in a counterclockwise orbit around the projectile (it sees an almost static electric field from the projectile at the timescale of the gyromotion). 
This leads to a situation where, even though the background particle starts to the right of the projectile, the $\vc{E}\times\vc{B}$ drift takes it to the left side of the projectile for a large fraction of the time interval over which the particles interact strongly.  
This results in momentum transfer in the $+\hat{y}$ direction when the projectile starts to the left of the background particle. 
In the opposing case, when the projectile starts to the right of the background particle, there is still a counterclockwise $\vc{E} \times \vc{B}$ drift as shown in the panels (d) and (h). 
However, the drift is not as pronounced in this case because it acts to shorten the interaction time between the particles by deflecting the background particle behind the projectile. 
For this reason, the background particle remains on the left side of the projectile for the entire interaction, and the net force is in the $+\hat{y}$ direction. 
Since the background particles starting from both the left and right sides of the projectile act to transfer momentum in the $+\hat{y}$ direction, the net gyro force is also in the $+\hat{y}$ direction (which is the $-\hat{n}$ direction).

 The critical speed at which the gyro friction changes sign can be estimated by comparing the interaction timescale ($\tau_c$) to the gyroperiod of the background particle $\omega_{c}^{-1}$. 
The interaction timescale is set by the ratio of the interaction length and the speed of the projectile: $\tau_c \sim l/v_1$. 
The interaction length ($l$) is approximately characterized by $\lambda_D$ in the weakly coupled regime, and $a$ in the strongly coupled regime. 
If $\tau_c \gtrsim \omega_{c}^{-1}$, then the background particle undergoes complete gyro-orbits during the collision and can thus exhibit $\vc{E} \times \vc{B}$ drifting motion. 
Note that gyration during a collision also requires that $r_c < l$, where $l$ is the interaction range, but we are already assuming that this condition is met ($\beta > 1$) and concentrating here on the dependence on the projectile speed. 
In this sense, it defines a speed-dependent parameter to characterize strong magnetization when $v_1 > v_T$. 
The condition $\tau_c > \omega_c^{-1}$ is satisfied when the projectile is sufficiently slow 
\begin{equation}
\frac{v_1}{v_T} \lesssim \frac{l \omega_c}{v_T} \approx \frac{l}{\lambda_D} \beta. 
\end{equation}
When this is satisfied, the $\vc{E}\times \vc{B}$ drift is expected to cause $F_n$ to be in the $-\hat{n}$ direction. 
In contrast, when $v_1/v_T \gtrsim l \omega_c/v_T \approx (l/\lambda_D) \beta$ the interaction time is too short for $\vc{E} \times \vc{B}$ motion to occur, and we expect the net gyro force to be in the $+\hat{n}$ direction. 
The expectation that this transition speed increases proportionally with $\beta$ is consistent with the trend observed in panel (c) of Fig.~\ref{fig:betas_GCO}. 
Considering strong coupling effects, the interaction range scales as $l/\lambda_D \approx a/\lambda_D = \sqrt{3} \Gamma^{1/2}$. 
The prediction that the critical speed increases proportionally to $\Gamma^{1/2}$ in the strongly coupled regime also appears consistent with the data shown in panel (c) of Figs.~\ref{fig:Polar_Gamma_1}--\ref{fig:Polar_Gamma_100}.



\begin{figure*} [!htb] 
\centerline{\includegraphics[width = 7.5in]{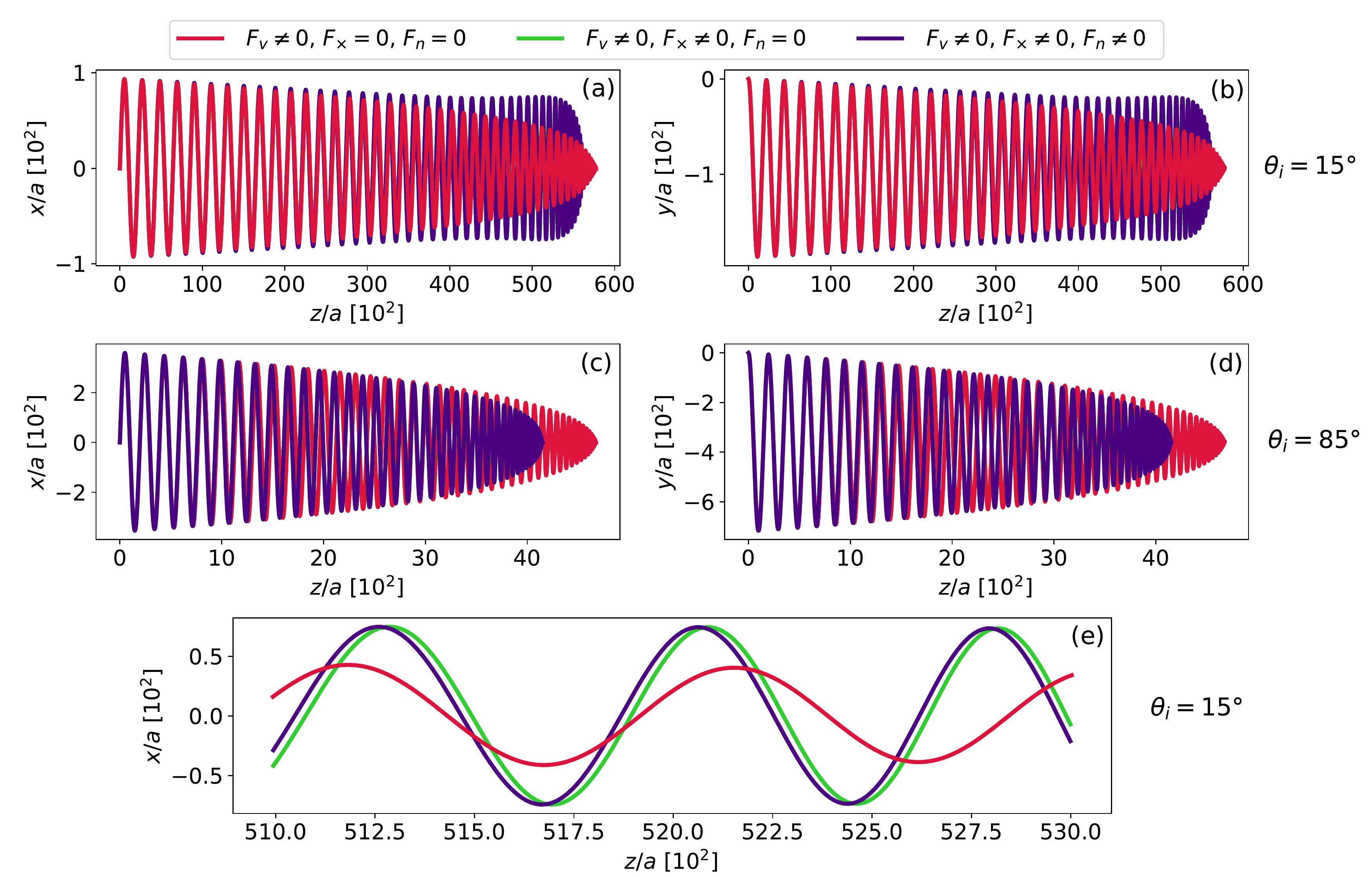}}
\caption { Trajectories of a massive projectile ($m_r =1000$) with initial speed $14 v_T$ moving through a one component plasma with coupling strength $\Gamma =10$ and magnetization strength $\beta=10$.}
  \label{fig:Trajectories}
\end{figure*}

\section{Trajectories}

The average motion of a projectile in a plasma is influenced by the Lorentz force and the friction force. Thus the equations of motion can be written as 
\begin{equation}\label{macroprojectile}
 m_t\frac{d \vc{v}}{dt} = q (\vc{v} \times \vc{B}) + \vc{F}.
 \end{equation}
Although the trajectory of an individual particle is influenced by diffusive motion resulting from interactions with other particles, Eq. (\ref{macroprojectile}) describes the expected trajectory resulting from the average of many sample individual trajectories. In order to study the effects of the various friction force components on the motion of a projectile through the plasma, we solve Eq. (\ref{macroprojectile}) including each of the three vector components of $\vc{F}$ shown in the polar plot figure \ref{fig:Polar_Gamma_10}. Figure \ref{fig:Trajectories} shows the resulting trajectory of a projectile moving in a background plasma characterized by $\Gamma =10$ and $\beta = 10$. The projectile starts at the origin with an initial speed of $ v =14v_T$. The initial velocity is $v_z = v \cos \theta_i $, $v_x = v \sin \theta_i $, $v_y = 0$ and initial orientation with respect to the magnetic field is $\theta_i = 15 \degree$ (panel (a), (b) and (e)) and $\theta_i = 85 \degree$ (panel (c) and (d)). Here the magnetic field direction is taken along the $\hat{z}$ direction. The trajectories are calculated for three different scenarios to emphasize the importance of different friction components: 1) includes the stopping power ($F_v \neq 0$), but excludes the transverse and gyro friction components ($F_\times = 0$ and $F_n = 0$) (red line). 2) includes the stopping power ($F_v\neq 0$) and transverse force ($F_{\times}\neq 0$), but excludes the gyro friction component ($F_n = 0$) (green line). 3) includes all the three components ($F_v \neq 0$, $F_{\times} \neq 0$ and $F_{n} \neq 0$) (blue line). The green and blue lines are only separated by a small phase shift and appear overlapping in panels (a) -- (d). Panel~(e) zooms into the phase shift. 

Out of the three components of the friction force, only stopping power decreases the kinetic energy of the projectile. It acts opposite to the motion of the projectile and reduces both the parallel and perpendicular kinetic energy. This results in a monotonic decrease in the gyroradius of the projectile (red line in Fig. \ref{fig:Trajectories}). The transverse force is perpendicular to the velocity of projectile and does not influence its total kinetic energy. However, it influences the gyroradius by redirecting parallel speed to perpendicular speed, or vice-versa (green line). For larger speeds, the transverse force is positive and it increases the gyroradius, whereas for smaller speeds, it is negative and decreases the gyroradius. This effect is prominent in the case of $\theta_i = 15\degree$. Here, the gyroradius increased for most of the motion and steeply decreased near the stopping point after the projectile speed had dropped sufficiently that the transverse force changed sign.

The transverse force also changes the stopping distance of the projectile. It is not a prominent effect at $\theta_i = 15^\circ$, but is well demonstrated by the $\theta_i = 85 \degree $ initial condition. In this scenario, the initial parallel speed is very low compared to previous scenario leading to a smaller stopping distance. But the transverse force strongly alters its trajectory by decreasing the parallel speed, resulting in a significantly shorter stopping distance.  

The gyro friction force is in the direction of the Lorentz force and is observed to create a small phase shift in the motion of the projectile; see panel (e) in Fig.~\ref{fig:Trajectories}. For larger speeds, the gyro friction is positive and it increases the gyrofrequency and for smaller speeds it is negative and decreases the gyrofrequency. The effect of gyro friction in the overall evolution of the projectile is not as prominent as the transverse force or stopping power. This can be deducted by writing the equations of motion using the  scaled variables in spherical polar coordinates. On scaling the velocity using $v_T$, friction by $T/a$ and time using $\omega_p$, we get 
\begin{subequations}
\begin{eqnarray}
\frac{d\tilde{v}}{d\tilde{t}} &=& \frac{\tilde{F}_v}{\sqrt{6 \Gamma} m_r}, \\
\frac{d \theta}{d\tilde{t}} &=& \frac{\tilde{F}_\times}{\sqrt{6 \Gamma} m_r \tilde{v}}, \\
\frac{d \phi}{d\tilde{t}} &=& -\frac{\beta}{m_r} -\frac{\tilde{F}_n}{\sqrt{6 \Gamma} m_r \tilde{v} \sin \theta}. \label{phiterm}
\end{eqnarray}
\end{subequations}
Here, the variables with tilde (\~{}) on top represents scaled variables, $\theta$ is the polar angle which is same as the orientation of the projectile's velocity with the direction of the magnetic field and $\phi$ is the azimuthal angle. On comparing the magnitude of the terms in the Eq. (\ref{phiterm}), the $\tilde{F}_n$ term is $10^3$ - $10^4$ times smaller than the $\beta$ term. On the other hand, other friction force components do not have any external force to compete with in the equations of motion. This explains the small effect of the gyro friction component in the trajectories of the projectile compared to other two components.

The overall movement of the projectile inside the plasma is primarily determined by the combined effect of both the stopping power and the transverse force. The effect of the gyro friction is comparatively weak because it competes with the large Lorentz force. However the gyro friction may influence other macroscopic transport or wave properties by changing the gyration rates.

\section{Conclusion}

This work has extended the generalized Boltzmann kinetic theory to the strongly coupled regime. The utility of the collision operator was shown by calculating the friction force on a test charge moving through a background plasma at conditions that range from weakly to strongly coupled, and weakly to strongly magnetized. Good agreement was found between the results from the GCO calculation and previous MD simulations. The combination of strong coupling and strong magnetization was found to introduce a third "gyro friction" component in the direction of the Lorentz force. This is in addition to the stopping power and the transverse friction force components previously studied in the weakly coupled, strongly magnetized regime~\cite{lafleur2019transverse,lafleur2020friction,Jose_POP_GCO}. Similar to the transverse force, the magnitude and the sign of the gyro friction depends on the speed and the orientation of the projectile's velocity with the direction of the magnetic field. It was found to be zero when the projectile moves parallel or anti-parallel to the magnetic field and have a maximum magnitude when the projectile's velocity is perpendicular to the magnetic field.
 
 The gyro friction arises due to asymmetries associated with gyromotion of background particles during close collisions, as well as an $\vc{E} \times \vc{B}$ drift. A competition between these effects was found to explain the sign reversal of the force with respect to the projectile speed. The trajectory calculation of the projectile showed that the transverse force and the gyro friction influences the overall evolution of the projectile. The transverse force changes the gyroradius and alters the stopping distance and the gyro friction slightly modifies the gyrofrequncy resulting in a phase shift. The effect of the gyro friction on the trajectory of the projectile was found to be small due to the large Lorentz force term. However, the phase shift that it causes may influence some transport processes in comparison to its absence.  Future works could explore this.
 
This work, along with the previous work~\cite{Jose_POP_GCO}, has shown the effectiveness of generalized collision operator to calculate the friction force in all the transport regimes of the magnetization-coupling phase space. Calculation of the friction force in these regimes has shown new physical effects - the transverse friction force and the gyro friction force. Other transport properties like electrical resistivity, self diffusion or thermal relaxation that are linked to the friction force are expected to show analogous qualitative changes with strong magnetization.  How the strong magnetization modifies these transport properties are yet to be determined. Answering this question would benefit many magnetized experiments such as non-neutral plasmas, ultra-cold neutral plasmas and magnetized dusty plasmas. The generalized collision operator is a strong candidate to explore these novel states of plasma.   

\section{Data Availability Statement}

The data that support the findings of this study are available from the corresponding author upon reasonable request.
 
\begin{acknowledgments}
The authors thank David J. Bernstein for helpful conversations during the development of this work.
This material is based upon work supported by the U.S. Department of Energy, Office of Fusion Energy Sciences, under Award No. DE-SC0016159, and the National Science Foundation under Grant No. PHY-1453736. It used the Extreme Science and Engineering Discovery Environment (XSEDE)~\cite{XSEDE}, which is supported by NSF Grant No. ACI-1548562, under Project Award No. PHYS-150018.
\end{acknowledgments}

\bibliography{references}	

\end{document}